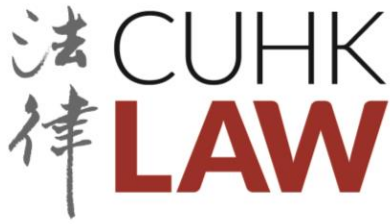

THE CHINESE UNIVERSITY OF HONG KONG

FACULTY OF LAW

Research Paper No. 2019-13

# Blockchain of Things (BCoT): The Fusion of Blockchain and IoT Technologies

Mahdi H. Miraz

# Chapter 07
# Blockchain of Things (BCoT): The Fusion of Blockchain and IoT Technologies

*Mahdi H. Miraz*

## 7.1 Introduction

Both Blockchain and Internet of Things (IoT) are the two major disruptive emerging constituents of the contemporary internet-enabled era of technology. As per Gartner Hype Cycle of Emerging Technologies 2018 (Panetta, 2018), both of these technologies are currently in their "peak of inflated expectations" while both are projected to highly likely require another "5 to 10 years" to mature. In fact, comparing with the Gartner's 2017 (Gartner, 2017) predictions, Blockchain - without changing much - hovered at its current ongoing position on the hype cycle. On the contrary, the locus of IoT has progressed reasonably – prevailing within the same arc (i.e. peak of inflated expectations) of the curve – moving downwards crossing the *pinnacle* – however, IoT pedalled back on the level of maturity from "2 to 5 years" to the current state of "5 to 10 years". Such regression of IoT, in terms of reaching maturity level, however, is justified by its widespread adoption in multifaceted applications and the security concerns raised thus far. In fact, both of these technologies are distributed, autonomous and *mostly* decentralised systems possessing connatural potentials to act as complementary to each other. IoT requires strengthening its security features while Blockchain inherently possesses them due to its extensive use of cryptographic mechanisms and Blockchain – in an inverted manner – needs contributions from the distributed nodes for its P2P (Peer-to-peer) consensus model while IoT rudimentarily embodies them within its architecture. This chapter, therefore, acutely dissects the viability, along with prospective challenges, of incorporating Blockchain with IoT technologies – inducing the notion of Blockchain of Things (BCoT) – as well as the benefits such consolidation can offer.

7.1.1 Introduction to Blockchain

The concept of Blockchain was first fully conceived as enabling technology for Bitcoin cryptosystem, as introduced in 2008 by a mysterious character called Satoshi Nakamoto (Nakamoto, 2008). However, expeditiously – within a very short span of time – Blockchain, for its wide possibility to be applied in multifaceted applications, has significantly proved its distinctiveness as a standalone technology. In fact, it can be argued that the Blockchain itself is not a new technology; it is rather a new concept of using different existing technologies in an incorporated approach (Miraz & Donald, Application of Blockchain in Booking and Registration Systems of Securities Exchanges, 16-17 August 2018).

Blockchain is a type of *Distributed Ledger (Also known as Shared Ledger or Distributed Ledger Technology, DLT)* – a shared database chronologically recording *transactions* – literally any sort and form of data – in a temper-proof digital ledger with time stamp. Blockchain ecosystem significantly utilises mathematical hashing and cryptographic asymmetric key encryption mechanisms for data security – along with P2P node based consensus approach for immutability. A brief operational description of Blockchain ecosystem has been presented at section 7.2.

7.1.2 Introduction to IoT

The phrase "The Internet of Things", more commonly known as "IoT", was first reportedly coined by one of the co-founders of MIT's Auto-ID Lab, namely "Kevin Ashton" far back in 1999. The term "Internet of Objects" is often used interchangeably. IoT ecosystem connects myriad of "things" or "objects" i.e. electronic or electrical devices- of different types, size, capabilities and characteristics- through the Internet. The principal aim is to maximise the benefits of data- in terms of practical usefulness as well as monetary gains by analysing and utilising in decision-making process - collected by various sensors and/or actuators embedded in different physical objects including machines. Major share of connectivity in any IoT ecosystem is mainly facilitated by a number of short-range wireless technologies such as: ZigBee, Radio Frequency Identification (RFID), Ultra-wideband (UWB) radio technology, sensor networks and through location-based technologies (Feki, Kawsar, Boussard, & Trappeniers, 2013). In fact, the latitude of such connections is continually extending beyond the scope of basic machine-to-machine (M2M) communication (Benattia & Ali, 2008). There are multifarious IoT devices available. Examples include: Smart toys, Wearables (e.g. Smart watches, glasses etc.), Smart appliances (such as Smart TVs, Smart speakers, smart Bulbs), Smart meters such as thermostats, Commercial security systems and smart city technologies (such as those used to monitor traffic and weather conditions). IoT applications are also multifarious in nature. Many IoT ecosystems, performing various different tasks, have been developed thus far. Examples of such IoT enabled systems include: Nest Smart Home, DHL's IoT Tracking and Monitoring System, CISCO's 'Planetary Skin' - a global "nervous system", Smart Grid and Intelligent Vehicles, Smart Firms, Smart Schools and so forth. In fact, the scope of IoT applications have always been expanding since it was first implemented. Figure 7.1 demonstrates how various heterogeneous networks can be connected through IoT – a "network of networks" (Evans, The Internet of Things: How the Next Evolution of the Internet Is Changing Everything, 2011) making the Internet even pervasive.

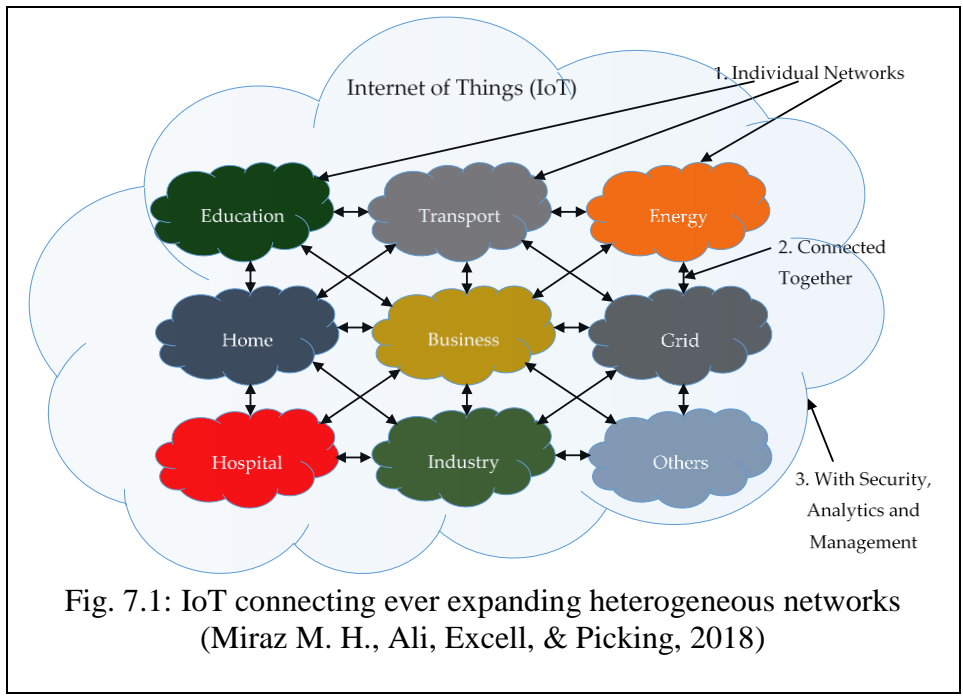

Fig. 7.1: IoT connecting ever expanding heterogeneous networks
(Miraz M. H., Ali, Excell, & Picking, 2018)

Apart from the wide range of standard networking protocols, domains and applications (Höller, et al., 10 Apr 2014) deployed in IoT ecosystems, IoT devices suffer from lack of standardisation, especially in terms of how they are connected to the Internet. However, this inhibiting factor is expected to be addressed in the near future.

7.1.3 Application of Blockchain in IoT
Blockchain and IoT – as standalone technologies – have already proved them to be highly disruptive.

Since IoT highly utilises the existing wireless sensor network (wsn) technologies, intrinsically it remains vulnerable to privacy as well as security threats. On the contrary, blockchain, by its design and architecture- consensus method and cryptographic techniques – is considered as a Trust Machine (Panarello, Tapas, Merlino, Longo, & Puliafito, 2018). Thus, it possesses the potentials to address major share of the security issues found in IoT. Miraz (Miraz & Ali, Blockchain Enabled Enhanced IoT Ecosystem Security, 2018) argues them to be complementary technologies to each other: BC requires participating nodes for consensus approach which can be supplemented by IoT devices while IoT requires security features which can be met by BC such as transparency, privacy, immutability, operational resilience and so forth.

IoT is a cyber-physical system which help represent the "connected" physical world into part of a substantial realm of information system – the cyber world. However, due to various reasons, the security aspects of IoT has not been properly addressed at the design phase of the devices and products. With the advent and increasing popularity of BC, there has been a paradigm shift in IoT research, particularly integrating IoT and BC (Samaniego & Deters, 2016; Sun, Yan, & Zhang, 2016) together for a more robust but secure cyber world. However, since the technologies are still not fully mature, many challenges are yet to be addressed as emerged from such integration (Reyna, Martín, Chen, Soler, & Díaz, 2018). Many studies (Jesus, Chicarino, Albuquerque, & Rocha, 2018; Kouicem, Bouabdallah, & Lakhlef, 2018) suggest applications of BC as a probable solution to tightening the security aspects of IoT ecosystem including the presentation of "Stalker" (Jesus, Chicarino, Albuquerque, & Rocha, 2018) attack.

Since IoT is built on the foundation laid by wireless sensor network (WSN) (Daia, Ramadan, & Fayek, 2018), characteristically each node of an IoT ecosystem is considered to be prone to attacks such as Distributed Denial-of-Service (DDoS) (Chaudhry, Saleem, Haskell-Dowland, & Miraz, 2018; Onik, Al-Zaben, Hoo, & Kim, 2018) and if compromised may serve as a point of failure.
IoT networks are mostly leveraged on cloud environment. Such centralised architecture suffers from Single Point of Failure (SPF) and further adds to vulnerability.

IoT devices gather and/or generate vast amount of data which are communicated over the Internet for processing and decision making purposes. Data privacy and authentication is considered to be a constant critical threat for IoT environment. In the absence of proper security measures, these vast amount of data can be mishandled and used inappropriately (Sicari, Rizzardi, Cappiello, Miorandi, & Coen-Porisini, 2017). It is thus extremely important to safeguard the IoT system from injection attacks. As the name implies, an injection attacks tries to inject false data or measures into the system and thus affect the overall decision making process.

### 7.1.4 Challenges in integrating Blockchain in IoT.

It is evident that the notion of Blockchain of Things (BCoT) – by creating a fusion of blockchain and IoT technologies, is capable of bring a paradigm shift in how these technologies are currently being used. Both the technologies can in fact benefit from each other in a reciprocal manner. However, integrating them together is not a straightforward matter. Many technological as well as architectural issues are yet to be solved for seamless integration. For instance, blockchain's Proof-of-Work (PoW) consensus approach may not be a good fit for IoT environment as it demands both computing power and electric energy to a great extent. Alternative approaches such as variants of Proof-of-Stake (PoS), Proof-of-Activity (PoA), Proof-of-Space/Capacity (PoC) are being designed, developed and implemented. Blockchain's capped latency and lower transaction throughput is another hurdle in its way to be applied in IoT environment. However, recent invention of Lightning Network (LN) and similar other technologies hold great promises to address this issue. Per contra, IoT highly devices highly suffer from scarce processing capabilities and lack of storage systems. In addition to the recent advancement in IoT devices, off shoring some processing and storage related functions to the cloud mitigate the problem to some extent. These challenges and status of recent developments in this regard, have been discussed in section 7.4 in more details.

## 7.2. Blockchain Fundamentals

### 7.2.1 Distributed Digital Ledger

In a blockchain ecosystem, there are mainly two types of nodes: full node and lightweight node. While the full nodes preserve the complete blockchain, lightweight nodes only download the headers of each block rather than the complete block. A lightweight node can also take part in the verification and consensus approach via connecting to a full node using simplified payment verification (SPV). Thus, downloading and storage requirement for a lightweight node is significantly reduced, however, this requires a lightweight node to place its "trust" on the associated full node instead.

Therefore, all (full) nodes are intrinsically complete ledgers - they hold and have access to data the whole blockchain data containing the complete transaction history in the chain. As stated in section 1.1, blockchain is thus seemingly a *Distributed Ledger (Also known as Shared Ledger or Distributed Ledger Technology, DLT)* – a shared but temper-proof digital ledger (database) of chronologically recorded time-stamped transactions or data. These transactions data, organised in blocks, are linked through the protocol along with hashing and consensus. Analogous to a ledger, an existing block cannot be deleted or modified as doing so will invalidate the "chain" of hashes. Like other ledgers, a DLT is append only – allowing to add new blocks at the open end of the chain by any participating or permissioned node. The process is controlled by the protocol via consensus approach without the need for any central authority.

One major advantage of this distributed approach is eliminating the single point of failure (SPF) as if one of the nodes becomes unavailable or compromised, the network shall still be functioning without any disruptions. Data are chronologically recorded in the ledger, thus it becomes easily verifiable. The decentralised approach, along with mathematical hashing provides immutability and transparency. DLT is also considered to be highly suitable for non-monetary transactions, especially for securities settlement. It is advocated that application of DLT can help bringing "direct" holding of securities and eliminate market fragmentation while bringing complete transparency in the settlement and clearing process (Donald & Miraz, 2019; Miraz & Donald, LApps: Technological, Legal and Market Potentials of Blockchain Lightning Network Applications, Forthcoming 2019).

7.2.2 Variations of Blockchain
Considering the permutation and combination of the *read* and *write* accesses assigned to the nodes, Blockchain ecosystems can be categorised into three different consensus models viz. public (permission less), private (permissioned) and hybrid (consortium).

Public (Permissionless):

In a public or permisionless blockchain ecosystem, anyone at any time and from anywhere in the world, having a computing device, can act as a participating node – joining and leaving the network at his or her own will. A node, willing to participate, has to install a small prototype which defines the consensus and other relevant rules. In most cases, all the nodes have both read and write access. However, nodes may opt out to be a "full node" – a node that keeps a copy of the "complete" ledger. Bitcoin's Blockchain is an example of public Blockchain ecosystem.

Private (Permissioned):
In a private or permissioned blockchain ecosystem, only "permitted" or "invited" nodes can be part of the network. These trusted nodes usually have both read and write access. However, a role based policy or even specific node based approach can also be applied. Multichain is an example of private blockchain.

Hybrid:
A hybrid blockchain, as the name implies, is a combination of both public and private models. While read access is usually left open for any participating nodes as in public blockchain, write access is rather confined to some specific nodes. Consensus is predominantly controlled by a group of predefined "trusted" nodes. Hybrid Blockchain can be considered as the best version of both the models, however, implementation decision should be based on the domains as well as the type of the applications. For example, hybrid blockchain may be a good choice for stock exchanges while public blockchain for cryptocurrencies (Miraz & Donald, Application of Blockchain in Booking and Registration Systems of Securities Exchanges, 16-17 August 2018).

Based on the history of evolution of this technology, Blockchain can further be categorised in four different versions thus far:

Blockchain 1.0
The type of blockchain behind Bitcoin cryptocurrency, as introduced by Shatoshi Nakamoto in 2008, is predominantly known as Blockchain 1.0. This sort of blockchain or DLT facilitates internet based financial transactions by enabling cryptocurrencies – the "Internet of Money".

Blockchain 2.0

As a rule of thumb, blockchains supporting smart contracts are largely known as Blockchain 2.0. Analogous to contracts, smart contracts – as coined by Nick Szabo in 1994 (Szabo, 1997) - are programmable digital contracts enabled by turing complete language. In its simplest form, smart contracts are autonomous computer programmes that can automatically execute if the predefined set of rules or conditions are met. These rules may include validation, verification, facilitation, administration of the execution of a contract and so forth. Vending machine is the oldest known example of materialising smart contract. In Blockchain 2.0, the smart contracts *reside* in the chain or DLT and thus inherit the built-in securities features that a blockchain can offer. Therefore, Blockchain 2.0 based smart contracts, along with security, offer variability and transparency. Ethereum Blockchain is the leading smart contrast enabled blockchain ecosystem.

Blockchain 3.0

Blockchain 3.0 supports the operation of Decentralised Applications (DApp), eliminating Single Point of Failure (SPF)- as seen in traditional centralised applications. DApps adopt decentralisation both in storage and communication aspects, therefore, the backend code of DApps are *mostly* run on blockchain ecosystems – decentralised peer-to-peer networks, while traditional apps utilises centralised servers to serve this purpose. Ethereum Swarm is an example of decentralised storage infrastructure allowing frontend code of DApps to host and run.

Blockchain 4.0

Blockchain 4.0 – based on the foundations already laid by its preceding variants – enables utilisation of the advent of blockchain technology in various applications, solutions, approaches and business models, especially in the realm of "industry 4.0" (cyber-physical systems). The pre-eminent driving force of Industry 4.0 is bringing complete automation in every phases of production systems (Onik, Miraz, & Kim, A Recruitment and Human Resource Management Technique Using Blockchain Technology for Industry 4.0, 2018). Such automation requires seamless integration of multifaceted execution systems as well as implementation of enterprise resource planning (ERP) – demanding highly reliable privacy protection and consensus model. This is where both IoT and blockchain kicks in – IoT providing the infrastructure for automation while blockchain acting as the "Trust Machine" [ (Miraz, Blockchain: Technology Fundamentals of the Trust Machine, 2017)]. Recent advent of atomic cross-chain swap and lighting network (Miraz & Donald, Atomic Cross-chain Swaps: Development, Trajectory and Potential of Non-monetary Digital Token Swap Facilities, 2019), is likely to accelerate the whole automation process of Industry 4.0 by a degree of great extent as it will enable swapping of IoT generated data on various Blockchain 4.0 applications on different platforms.

7.2.3 PoW vs PoS
As per the blockchain architecture, it is obvious that consensus approach is required to verify and validate transaction and then assemble them in a block for chaining with the existing ledger. The final step is basically sealing a newly built block incorporating some or all from a pool of verified but unconfirmed transactions. This process involves calculating the hash of the block for making it immutable and verifiable in the future. To avoid "double spending" of the same coin, the process requires "someone" to have the authority to seal a block for addition at some given point of time. There are various algorithms, such as Proof-of-Work (PoS) and Proof-of-Stake (PoS) to determine this "someone" by the protocol, rather than any central administrator.

PoW is the most commonly used algorithm, as ushered by Bitcoin. Bitcoin miners, who operate full bitcoin nodes, pull some transactions from the pool of unconfirmed transactions, add a new coinbase transaction to oneself to create a new coin as per the mining reward rate of that given time, add a nonce and then calculate the block hash. That being said, the hash has to solve the mathematical puzzle i.e. it has to smaller than a given threshold more commonly known as the "difficulty level". If the first hash, calculated by a particular miner, do not satisfy the difficulty level threshold, the nonce is changed, usually by adding one to it, and repeatedly calculated until the satisfying solution is found – similar to a brute force approach. Whoever finds the solution first, amongst all the miners, is the winner and receives the newly created coin. Thus, analogous to gold mining, the process of completing the PoW is known as mining too. Once the satisfying has is found, it is broadcasted to the network, other nodes then verify it and if found to be legitimate, the block is then added to their existing chain and they start working on forming a new block by repeating the same procedure. It is possible that two different miners produce the valid hashes at the same or nearly same time. In that case, the "longest chain" rule shall be applied to avoid any fork. The difficulty level is also automatically adjusted by the Bitcoin protocol to keep it approximately 10 minutes on average. The overall PoW consensus approach as well as this capped latency thus contribute to high latency which is one of the major impediments of blockchain adoption. High demand for computing power as well as electricity is another major issue for which PoW and mining is highly critiqued. Considering the limited computing power of IoT devices, PoW is not a good fit for the fusion of blockchain and IoT technologies - Blockchain of Things (BCoT), where IoT devices act as participating nodes.

Proof-of-Stake (PoS) is an alternative approach to PoW. In PoS, instead of solving cryptographic puzzle as part of mining competition, the amount of stake (wealth, cryptocurrency) a node possesses, incorporated with algorithms for randomisation, is considered while determining the creator of the next block. The more wealth/stake a node posses, the higher is the possibility for being selected as the creator of next block. While it is argued that PoS is more suitable, at least considering its currents state of development, for non-monetary applications of blockchain, DASH cryptocurrency has already adopted this approach and Ethereum has included in its future development roadmap. While PoS is considered to be less secure than PoW, it is more eco-friendly as it produces less Electronic Waste (E-Waste) and produces comparatively very less Green House Gas Emission (GHGE) by consumes less electricity (Miraz & Peter, Evaluation of Green Alternatives for Blockchain Proof-of-Work (PoW) Approach, 2019).

There are now few other emerging approaches, as alternatives to PoW, such as Proof-of-Activity (PoA), Proof-of-Burn (PoB), Proof-of-Capacity/Space (PoC) and Transactions as Proof of Stake (TaPoS). However, mostly all of these alternatives are prone to centralisation stands against the decentralisation notion of bitcoin i.e. to function as a "Trust Machine" through shifting the trust to a decentralised network from the third-party intermediaries.

7.2.4 Benefits of Blockchain
Based on Blockchain's architecture and functionalities, it is evident that blockchain offer the following benefits:
Decentralisation: The first and foremost benefits of blockchain it operates in a distributed network and the ledger is replicated in all the participating nodes. Therefore, all other benefits of blockchain is mainly derived from its decentralisation nature.
Transparency: Since transactions are recorded and timestamped in a decentralised ledger, blockchain transactions are completely transparent. Blockchain made verification of transaction further effortless through the application of Merkle Tree. Another important aspect of transparency is that the ledger can be precisely tracked back along the chain, authoritatively as well as accurately, to its point of origin.

Security: Since the ledger is distributed, SPF is eliminated. Furthermore, consensus approach, such as PoW, and longest chain rule makes the blockchain network protected from DDoS by capturing 51% or more nodes.

Immutability: Since all the time-stamped records of transactions are linked by mathematical hashing, altering one single transaction in the chain invalidates not only the hash of block it belongs to but also the hashes of all other blocks generated after that particular block. Per contra, replica of the chain is distributed on all the nodes of the network which provides verifiability – making the chain completely immutable. The ledger being append-only adds extra layer of immutability as existing record on the ledger can neither be deleted nor altered.

Cost: For large scale applications, deploying blockchain could be well of legacy technologies and will need less maintenance – making blockchain an economical and affordable solution in the long run. On small-scale private application, it may be expensive to deploy blockchain as it requires a distributed network to operate. However, various Blockchain as a Service (BaaS), quite a lot of which are even cloud based, offered by many third-party platforms, such as Ethereum, Hyperledger etc., can be utilised for offshoring purposes.

Smart Contracts: As discussed in section 7.2.1, smart contracts and Decentralised Applications (DApp) are now acting as a catalyst for blockchain adoption in various domains, including non-monetary ones. Smart contract enables pre-setting conditions on the blockchain. If the predefined condition or set of conditions are met, the blockchain system automatically triggers the transactions or materialise the contracts.

Lightning Network:
In a Lightning Network (LN), Hashed Timelock Contract (HTLC) based smart contract enables direct bi-directional transactions to take place between two parties. The intermediate transactions in LN network takes place in a second layer – built on top of the base layer of any blockchain ecosystem. These transactions are not subject to consensus, hence instantaneous. However, upon leaving the LN, the final resultant balance is broadcasted to the base layer network for consensus and settlement. Utilising onion-style routing, the scope of transactions in lighting networks can be expanded beyond directly connected peers.

Micropayment:
Drivers of Future Business Models: With the advent of smart contract and lightning network supported by blockchain, Lightning Applications (LApps) and truly affordable micropayment systems have emerged. These are now acting as drivers of future business models – by prompting innovation and nurturing new venture creations.

### 7.3  IoT Fundamentals
7.3.1 Internet of Everything, Things and Nano-things (IoE, IoT and IoNT)

It has been observed that the terms Internet of Everything (IoE) and Internet of Things (IoT) are often inappropriately used interchangeably. This is thus very important to distinguish between IoE and IoT. In fact, both Qualcomm and Cisco have been using the term IoE (Weissberger, 2014; Evans, 2012). However, while Qualcomm's connotation of IoE has been overridden by IoT by a majority of others, Cisco's interpretation is much more comprehensive. Cisco definitions of IoE comprises of "four pillars": people, data, process and things, where "things" characterises IoT (Miraz M. H., Ali, Excell, & Rich, 2015), refer to the figure 7.2 below.

With the advent of modern communication technologies, the detached, non-networked and solitary multifaceted devices of the past are now increasingly being connected through the Internet, including person-to-person (P2P) systems, person-to-machine (P2M) and even machine-to-machine (M2M) connectivity. The complete notion of IoE thus envelops people, processes, data and things together – in a close association, as shown in figure 7.2. Consequently, IoE innately supplements, enhances and broadens both industrial and business processes to enrich people's lives.

Further to the introduction of IoT in section 7.1.2, a conventional IoT ecosystem comprises of five distinct components – functioning in a way that involves mutual assistance in working towards a common goal. These components are as follows:
(1) Sensors: The sensors mainly functions as "input devices" which collect as well as transduce the data they sense;

(2) Computing Node: The computing node is mainly a processor – to further process the information and data received from a sensor;
(3) Receiver: The role of the receivers involves collecting the message or data sent by the computing nodes or any other associated devices;
(4) Actuator: The actuator is primarily responsible for triggering the associated device to perform the desired function as instructed by the computing node - based on its decision deduced by processing and analysing the data and information gathered from the sensors and/or the Internet;
(5) Device: The devices "actually" perform the desired task(s) as and when triggered by the actuators.

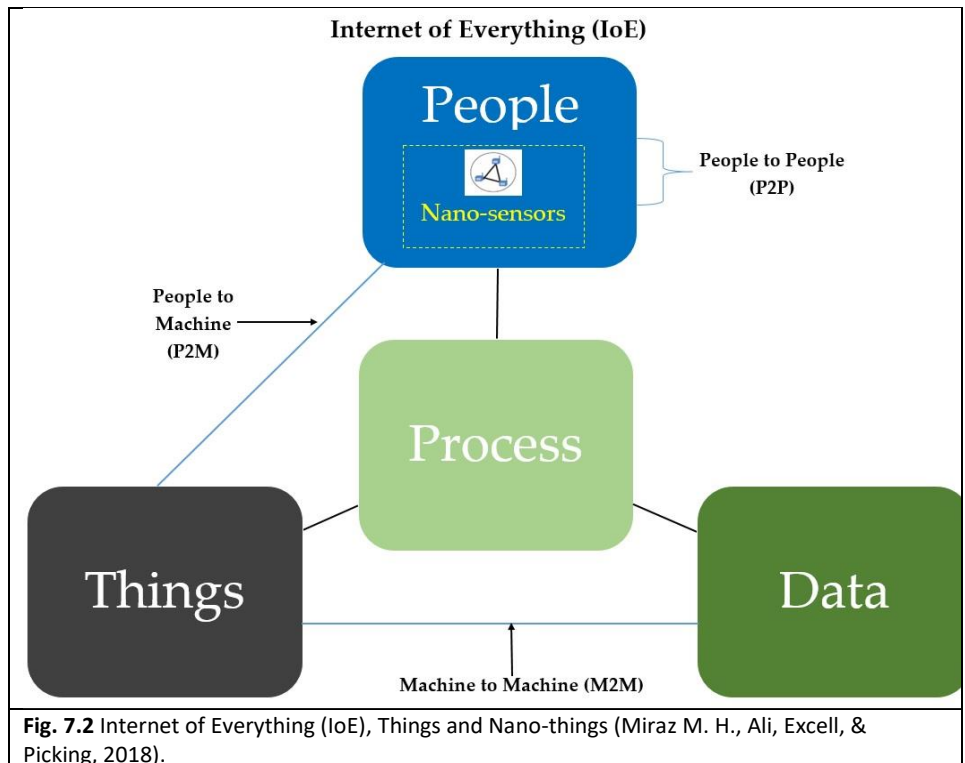

**Fig. 7.2** Internet of Everything (IoE), Things and Nano-things (Miraz M. H., Ali, Excell, & Picking, 2018).

The notion of Internet of Nano-things (IoNT) is fundamentally a genre of Internet of Things (IoT) – standard sensors being replaced by nano-sensors. IoNT embodies nano-sensors in multifaceted objects with the advent of nano-networks. IoNT possesses significant potentials to highly benefit healthcare sectors by enabling access to healthcare data from various in situ places of the human or animal body which were inaccessible in the past due to the comparatively "large" size of the sensors. In fact, in an IoNT ecosystem, the functional endeavours i.e. sensing or actuation is to be performed by a "nano-machine" of dimensions ranging from one to 100 nanometre (nm), utilising nano-antennas operating at Terahertz frequencies. It is likely that IoNT is due to bring better medical diagnostics (Balasubramaniam & Kangasharju, 2013). Considering the potentials of IoNT in the healthcare sector, the nano-sensors in figure 7.2 has been placed inside the "people" box even though the nano-sensors are mainly small-scale sensors.

### 7.3.2 Challenges of IoT

Due to architectural limitations, the major challenges of IoT includes sustainable source of energy, scarce processing capability and security. There are many other limitations of IoT which needs to be addressed are meticulously. Such challenges of IoT includes (but not limited to): deployment of IPv6, lack of standardization, pervasiveness of IoT applications as well as devices, retrofitting IoT devices with additional sensors, lack of scalability to meet multifaceted exponential growths, amalgamating with the software defined networks (SDN) paradigm, to meet the increasing demand for performance requirement of edge computing (fog), inherent limitations of current wireless sensor networks (WSNs), ethical and legal issues – especially those related to data ownership and data residency, identity management of connected devices while enabling automated discovery, meeting future database and data management stipulations (Miraz M. H., Ali, Excell, & Picking, 2018). In fact, it was difficult to predict the wide-spread adoption of IoT at the initial phase of development, therefore, not much attention has been given at the design phase of the devices. This ignorance has resulted in the huge challenges IoT is facing at this stage.

### 7.3.3 IoT Security

It is apparent that one of the primary challenges IoT has to overcome is the drawbacks associated with security, privacy and vulnerability aspects. IoT systems highly suffer from SPF vulnerability due to their cloud based centralised configuration. IoT also suffers from device authentication and data confidentiality. If proper security measures are not in place, IoT systems can be compromised and used inappropriately.

It is pertinent to protect IoT systems from any attacks such as DDoS and injection attacks. While DDoS aims to disrupt regular legitimate traffic of a targeted network, server or service, injection attacks aims to disrupt decision-making by injecting false measures in the data. Both availability and data integrity is extremely important for any real-time and life critical applications such as healthcare, vehicular networks etc. Thus creating trust amongst IoT devices is extremely important and considered as a significant challenge. However, application of blockchain can significantly improve IoT security in this regard. Blockchain can offer IoT the required mechanism to achieve publicly verifiable audit trail through (device) authentication and (data) hashing techniques used blockchain ecosystems. This can thus help solving the problem of non-repudiation to a great extent.

### 7.4 Application of Blockchain for Enhanced IoT Security

The benefits BC can offer, such as security, transparency, immutability, verifiability as well as the smart contract and the LN based ones, possess the capability the limitations of IoT ecosystem if combined together with BC. Per contra, IoT also possesses the capability to benefit BC by actively participating at the consensus process. In the Blockchain of Things (BCoT)- the fusion of BC and IoT technologies- both can benefit from each other in a reciprocal manner. This section will present a detailed literature survey covering wide range of projects and research on the integration of BC and IoT i.e. the BCoT notion.

In fact, due to the mushrooming popularity of both BC and IoT, many researchers around the globe are now trying to innovate different ways of BC-IoT integration for developing highly secure but robust Information Technology (IT) systems and addressing the technical as well as other associated problems. The works of Sun *et al.* (Sun, Yan, & Zhang, 2016), Samaniego & Deters (Samaniego & Deters, 2016) and Reyna et al. (Reyna, Martín, Chen, Soler, & Díaz, 2018; Atzori, Iera, & Morabito, 2010) are worth mentioning in this regard. Many studies (Jesus, Chicarino, Albuquerque, & Rocha, 2018; Kouicem, Bouabdallah, & Lakhlef, 2018) suggest applications of BC as a probable solution to tightening the security aspects of IoT ecosystem including the presentation of "Stalker" (Jesus, Chicarino, Albuquerque, & Rocha, 2018) attack.

Another research on studying the advantages and disadvantages of application of BC in IoT, by Christidis and Devetsikiotis (2016), introduces a taxonomy of BC topologies for this purpose. In fact, several divergent abstractions have been introduced with Proof-of-Concept (PoC) prototypes. Examples of such PoC include: application of blockchain together with InterPlanetary File System (IPFS) for upgrading firmware of IoT devices by utilising smart contracts, framework for generating cash flow by facilitating resource (LO3 Energy, 2017)or data (Protocol Labs, 2017) trading.

The Filament[1] research projects involve designing and developing a wireless network cable of controlling "any" system – ranging from street bulbs of a city to burglar alarm system of any office. That being said, the projects highly focus on the use of blockchain and smart contract to enable smart devices (such as sensors, smart refrigerator, smart TV or any other smart appliances) to interact with each other via seamless Machine-to-Machine (M2M) communications including discovering and exchanging messages – autonomously, without being controlled by any central authority. However, every any communication takes place, the devices have to authenticate themselves, by either Transport Layer Security (TLS) or Secure Socket Layer (SSL) protocols, for security purposes, could be using public key infrastructure (PKI). Another such M2M intercommunication model amongst smart IoT devices utilising blockchain as the backbone was proposed by Prabhu and Prabhu (2017). In this proposed model, the IP addresses of the devices as a key for accessing information stored in a DLT or blockchain.

With regard to access control, most of the established Access Control Lists (ACLs) and authentication approaches for traditional networks do not fit well in an IoT environment. This is mainly because of the centralised nature of ACLs and similar approaches such as Discretionary Access Control (DAC), Mandatory Access Control (MAC) and Attribute-Based Access Control (ABAC). To address these problems, a model was proposed by Deters (2017), to perform access control in an IoT environment utilising the statistics extracted from the access patterns, along with blockchain and smart contract.

---

[1] https://filament.com/

A multi-tier architecture of BCoT security and privacy model has been proposed by Dorri *et al*. (2016; 2017) eliminating the shortcomings of BC as well as other traditional approaches. Similar level of confidentiality and data integrity was achieved without the use of PoW. Their system is designed based on three different layers: smart home, overlay network and cloud storage. Apart from smart devices, the smart home also has a miner who governs the blockchain as well as the data access policies. When a new device (node) is added to the smart home ecosystem, the miner creates and adds a new block corresponding to the newly added node. In fact, the newly added block possesses dual-header- i.e. block header containing a link to the preceding block while and policy header defines the data access rule and authority. For facilitating secure communicate amongst the devices shared keys are used – created and distributed utilising Diffie–Hellman algorithm, governed by the miner. A smart device, in this system, can store the data either on local storage system by employing s shared key or on a cloud storage by sending a request to the miner will then trigger a transaction on the public blockchain – the transaction is signed with the device's key and contains addresses of the cloud storage system. Thus the proposed BCoT architecture provides fivefold security and privacy related benefits: 1) confidentiality through use of shared private key encryption, 2) integrity through hashing, 3) availability by limiting allowed transactions, 4) user control by blockchain technology and 5) authorisation by applying authorisation policies along with utilising shared key.

Wörner and Bomhard (2014) developed a BCoT system enabling network sensors to trade and exchange data for Bitcoins in a self-governing fashion. Nodes' addresses are same as public keys on Bitcoin network. Sensor nodes are discoverable via sensor repository. If a client would like to receive data from a sensor node, the client has to send transaction (including payment in Bitcoin) addressed to the public key of the corresponding sensor. The sensor node will then send a response transaction (including data) to the public key of the client. The delivery of the data in such scenarios can be processed through smart contracts. An alternative approach of using Bitcoin or similar altcoins could be using IOTA – a cryptocurrency using no blocks and no miners while facilitating micropayments (IOTA Foundation, 2018).

Chakraborty et al. (2018) has recently advocate a two-layered architecture to address the security as well as resource-constrained aspects of IoT nodes. The nodes having limited resource for enforcing security measures are clustered together in layer 0. Per contra, other primary and secondary nodes are congregated in level N – while the primary nodes takes care of the relevant processing, the secondary nodes mainly assist the primary nodes in this regard. The resource limitations of the nodes in layer 0 prevents them from communicating directly with other layer 0 devices, however, this rather achieved via level N devices instead.

## 7.5 References
(**References to be used in the Book Chapter**)

## 7.6  References for Advance/Further reading

for Computing Machinery (ACM). Retrieved from https://papers.mathyvanhoef.com/ccs2017.pdf

Vermesan, O., & Friess, P. (2 July 2013). *Internet of Things: Converging Technologies for Smart Environments and Integrated Ecosystems* (1st ed.). Aalborg, Denmark: River Publishers. Retrieved June 18, 2018, from http://www.internet-of-things-research.eu/pdf/Converging_Technologies_for_Smart_Environments_and_Integrated_Ecosystems_IERC_Book_Open_Access_2013.pdf

Weber, R. H. (2010, January 1). Internet of Things – New security and privacy challenges. *Computer Law & Security Review, 26*(1), 23-30. doi:10.1016/j.clsr.2009.11.008